\documentclass[nofootinbib,showpacs,preprintnumbers,amsmath,amssymb]{revtex4}
\usepackage{graphicx}

\begin{document}


\title{Running coupling effects for the singlet structure function $g_1$
at small $x$}

\vspace*{0.3 cm}

\author{B.I.~Ermolaev}

\altaffiliation[Permanent Address: ]{Ioffe Physico-Technical Institute, 194021
 St.Petersburg, Russia}
\affiliation{CFTC, University of Lisbon,
Av. Prof. Gama Pinto 2, P-1649-003 Lisbon, Portugal}
\author{M.~Greco}
\affiliation{Dept of Physics and INFN, University Rome III, Rome, Italy}
\author{S.I.~Troyan}
\affiliation{St.Petersburg Institute of Nuclear Physics,
188300 Gatchina, Russia}

\begin{abstract}
The running of the QCD coupling is incorporated into the infrared
evolution equations
for the flavour structure function $g_1$.
The explicit expressions for $g_1$ including the total resummation of the
double-logarithmic contributions and accounting for the running coupling
are obtained. We predict  that 
asymptotically $g_1 \sim x^{- \Delta_S}$, with the intercept
$\Delta_S = 0.86$, which is more than twice larger than the non-singlet
intercept $\Delta_{NS}= 0.4$. The impact of the initial quark $(\delta q)$  
and gluon  $(\delta g)$  
densities  on the sign of $g_1$ at $x \ll 1$ is discussed 
and explicit expressions 
relating  $\delta q$ and $\delta g$ are obtained. 
 
\end{abstract}


\maketitle

\section{Introduction}

The flavour singlet structure function $g_1$
has been the object of intensive theoretical investigations. First $g_1$ 
was
calculated in Refs.~\cite{ar},\cite{ap}, where the LO DGLAP evolution
equations\cite{ap},\cite{dgl} were used.
Since that, the DGLAP approach has become
the standard instrument for the theoretical description of  the
singlet and non-singlet components of $g_1$ and provides
 good agreement between experimental and theoretical results (see e.g.
Ref.~\cite{a} and the recent review \cite{b}). On the other hand, despite
this agreement, it is known that from a theoretical point of view
the DGLAP equations are not supposed to work
well at $x \ll 1$ because the expressions for the anomalous dimensions 
and the coefficient functions in 
these equations account only for a finite part of the NLO
$(\sim 1/ \omega^{n + 1})$, with $n = 1,2$ (see \cite{grsv}).
In particular, this means that all
double-logarithmic (DL) contributions
$\sim (\alpha_s \ln^2 (1/x))^n, ~n > 1$ are neglected in the DGLAP
expressions for $g_1$.
As a matter of fact, these contributions  become very
important in the small $x$ region and should be
accounted for to all orders in $\alpha_s$. Such total resummation
 was done in  Refs.~\cite{ber1} for the non-singlet component, $g_1^{NS}$, 
of $g_1$ and in Ref.~\cite{ber} for the singlet one. These calculations
were done in the double-logarithmic
approximation (DLA) and it was shown that $g_1$ has the power-like
(Regge) behaviour $\sim x^{- \Delta}$
when $x \to 0$. However, the QCD coupling
in Refs.~\cite{ber1,ber}
is kept fixed, whereas, as well known, the running
coupling effects are relevant.
As the DL intercepts $\Delta$ of $g_1$ obtained in \cite{ber1,ber}
are proportional to
$\sqrt{\alpha_s}$ fixed at an unknown scale, it makes the
results of Refs.~\cite{ber1,ber} unclear and not suitable for practical
use. A parametrisation for fixed
$\alpha_s = \alpha_s(Q^2)$ for  both
the singlet  and non-singlet component
was suggested in Refs~\cite{kw}- \cite{z}.
In the DGLAP framework,
 in a general ladder rung $\alpha_s$ depends on the transverse
momentum $k_{\perp}$ of the ladder parton (a quark or a gluon), with
$ \mu^2 < k^2_{\perp} < Q^2$, where $\mu^2$ is the starting point of the
$Q^2$ -evolution. The arguments in favour of
such a dependence were
given in Ref.~\cite{abcmv}.
However, in   Ref.~\cite{egt1} it was
shown that the arguments of Ref.~\cite{abcmv} are valid only when
$x$ is large ($x \sim 1$), while on the contrary
in the small-$x$ region, in a general ladder rung $i$
$\alpha_s$ depends
on the virtuality of the ``horizontal'' gluon,  $(k_i - k_{i-1})^2 $.
This dependence was used in Refs.~\cite{egt2}
for studying
$g_1^{NS}$ at small $x$, with running $\alpha_s$ accounted for. Our
prediction for the non-singlet
intercept was confirmed by phenomenological analyses
of the experimental data in Refs.~\cite{kat}.

In the present paper we apply the same arguments of  Ref.~\cite{egt1} in
order to account for the running $\alpha_s$ effects for the description of
the small-$x$ behaviour of the flavour singlet component of $g_1$.
The paper is organised as follows: In Sect.~2 we construct and solve
the system of the infrared evolution eqs. for  $g_1$. In Sect.~3  we
obtain the explicit expressions for
the anomalous dimensions of $g_1$. These expressions account for all DL
contributions and  also for running $\alpha_s$.
The intercept of $g_1$ is calculated in Sect.~4. 
In Sect.~5, we specify the general solutions obtained in Sect.~2 through
the phenomenological quark and gluon inputs and estimate the sign of $g_1$
at $x \ll 1$. Then we calculate the perturbative contributions to these
inputs.
Finally, Sect.~6 contains our
concluding remarks.

\section{IREE for the structure function $g_1$}

In order to obtain $g_1$ with all DL contributions accounted for at small
 $x$, we cannot use the DGLAP equations. Instead of that,
we construct for it a set of infrared evolution equations (IREE), i.e.
equations for the evolution  with respect to the infrared cut-off $\mu$ in
the transverse
momentum space. The cut-off $\mu$ is introduced as the starting point of
the evolution with respect to both $Q^2$ and  $x$. In contrast to the
DGLAP equations where only $Q^2$ -evolution is studied, the IREE are
two-dimensional.
In order to derive these IREE, it is convenient to operate with
the spin-dependent
invariant amplitude $M_q$ which can be extracted from the
forward Compton scattering amplitude
$M_{\mu \nu}$ by using the projection operator
$\imath \epsilon_{\mu \nu \lambda \rho} q_{\lambda} p_{\rho}/pq$.
Through this paper we use the standard
notations, so that $q_{\lambda}$ is the momentum of the off-shell photon
$(-q^2 = Q^2 \geq \mu^2)$ and $p_{\mu}$ is the momentum of the (nearly)
on-shell quark.
When $M_q(s, Q^2)$ is obtained, the polarized quark contribution to
$g_1$,  $g_q$, can be
easily found:

\begin{equation}
\label{g1m}
g_q(x, Q^2) = \frac{1}{\pi} \Im_s M_q(s, Q^2)
\end{equation}
where $s$ is the standard Mandelstam variable, $s \approx 2pq$.
The subscript $q$ in the rhs of Eq.~(\ref{g1m}) means that the off-shell 
photon is scattered by the quark.
It turns out that
the IREE for $M_q(x, Q^2)$ involve the spin-dependent
invariant amplitude, $M_g(s, Q^2)$ of the forward Compton scattering
where the off-shell photon is scattered by a nearly on-shell
gluon with momentum $p$. Similarly to Eq.~(\ref{g1m}), the polarized
gluon contribution, $g_g$ is related to $M_g$ as follows:
\begin{equation}
\label{g1primem}
g_g(x, Q^2) = \frac{1}{\pi} \Im_s M_g(s, Q^2) .
\end{equation}

Therefore, in our notations
\begin{equation}
\label{g1tot}
g_1(x, Q^2) = g_q(x, Q^2) + g_g(x, Q^2) .
\end{equation}
There is no difference between our
approach and DGLAP in this respect.
The main difference between the IREE  and
the DGLAP equations
is that due to the $k_{\perp}$-ordering which is the key point of
DGLAP, the ladder partons with
minimal transverse momenta should always be in the lowest
rung. However, such an assumption is an approximation which is
true only for large $x$. When $x$ is small, such partons
can be in any rung in the ladder.  More generally, each rung consists of
two ladder partons
which can be either quarks or gluons. Each of these two
cases yields DL contributions and therefore have to be accounted for.
It is convenient to write down the IREE in the $\omega$ -space related to
the momentum space through the asymptotic form of the Sommerfeld-Watson
(SW) transform for the scattering amplitudes. However, the SW transform is
defined for the signature amplitudes $M^{(\pm)}$. In particular for
the negative signature amplitudes it reads (we drop the superscript
``(-)'' as we do not discuss the positive signature amplitudes
in the present paper):

\begin{equation}
\label{mellin}
 M_r(s, Q^2) =
\int_{- \imath \infty}^{\imath \infty}\frac{d \omega}{2 \pi \imath}
(s/ \mu^2)^{\omega}
\xi(\omega)  F_r(\omega, Q^2)
\end{equation}
where $r = q, g$ and $\xi(\omega)$ is the negative signature factor,
$\xi(\omega) =  [1 - e^{ - \imath \pi \omega}]/ 2 \approx
 \imath \pi \omega / 2$.
It is necessary to note that  the transform inverse to Eq.~(\ref{mellin})
involves the imaginary parts of $M_r$:

\begin{equation}
\label{invmellin}
F_r(\omega, Q^2) =  \frac{2}{\pi \omega}\int_0^{\infty} d \rho
e^{- \rho \omega}
\Im M_r(s, Q^2)
\end{equation}
where we have
used the logarithmic variable $\rho = \ln(s/\mu^2)$. We will also
use another logarithmic variable $y = \ln(Q^2/\mu^2)$.

In these terms, the system of IREE for  $F_r(\omega, Q^2)$
can be written down as follows:

\begin{eqnarray}
\label{system}
\big( \omega + \frac{\partial}{\partial y}\big) F_q(\omega, y) =
\frac{1}{8 \pi^2} \big[ F_{qq}(\omega) F_q(\omega, y)  +
F_{qg}(\omega) F_g (\omega, y)\big] ~, \nonumber \\
\big( \omega + \frac{\partial}{\partial y}\big) F_g(\omega, y) =
\frac{1}{8 \pi^2} \big[ F_{gq}(\omega) F_q(\omega, y)  + F_{gg}(\omega)
 F_g(\omega, y) \big] ~
\end{eqnarray}
where the anomalous dimensions $F_{ik}$, with $i,k = q,g$,
correspond to the forward amplitudes
for quark and/or gluon scattering, having used the standard
DGLAP notations for the subscripts. It is convenient to absorb factors
$1/ 8 \pi^2$ in Eqs.~(\ref{system}) into the definition of new amplitudes
$H_{ik}$ related to $F_{ik}$ by

\begin{equation}
\label{h}
H_{ik}(\omega) = (1/ 8 \pi^2)F_{ik}(\omega) ~.
\end{equation}

The lhs of Eqs.~(\ref{system}) corresponds to apply the operator
$-\mu^2 (\partial/\partial \mu^2)$ to $M_k$. The first (second)
term in the rhs
of each equation ~(\ref{system}) corresponds to the case when the
ladder rung with minimal transverse momentum is made of a quark (gluon)
pair.
The Born contribution
does not appear in the rhs of Eqs.~(\ref{system}) because
the Born amplitude $M_g^{Born} = 0$ and
$M_q^{Born} = e^2_q s/ (s - Q^2 + \imath \epsilon)$ and therefore
it vanishes when differentiating with respect to $\mu$.
The main difference between the system of the
Altarelli-Parisi (AP) eqs. for the singlet
$g_1$ and Eqs.~(\ref{system}) is
that in the AP eqs. the amplitudes $H_{ik}$ contain
NLO terms with two-loop accuracy, whereas
$H_{ik}$ in Eqs.~(\ref{system}) account for
all NLO terms in DLA.
Then, these NLO terms cannot be included within the DGLAP approach
whereas in  our approach
we take them into account in Sect.~3, using   $\mu^2$
-evolution.
Solving Eqs.~(\ref{system}) and using Eqs.~(\ref{g1m},\ref{g1primem}),
we arrive at the following expressions for $g_q$ and $g_g$:
\begin{equation}
\label{g1general}
g_q(x, Q^2) = \int_{- \imath \infty}^{\imath \infty}
\frac{d \omega}{2 \pi \imath} (1/ x)^{\omega}
\Big[C_+(\omega) e^{\Omega_+ y} +
C_-(\omega) e^{\Omega_- y} \Big] ~,
\end{equation}

\begin{equation}
\label{g1primegeneral}
g_g(x, Q^2) = \int_{- \imath \infty}^{\imath \infty}
\frac{d \omega}{2 \pi \imath} (1/ x)^{\omega}
\Big[C_+(\omega) \frac{X + \sqrt{R}}{2H_{qg}} e^{\Omega_+ y} +
C_-(\omega) \frac{X - \sqrt{R}}{2H_{qg}} e^{\Omega_- y} \Big]
\end{equation}
Factors $C_{\pm}(\omega)$ will be specified in Sect. IV.
We have denoted here

\begin{equation}
\label{t}
X = H_{gg} - H_{qq} ,
\end{equation}

\begin{equation}
\label{r}
R = (H_{gg} - H_{qq})^2 +4 H_{qg}H_{gq} ~.
\end{equation}
and

\begin{equation}
\label{omega}
\Omega_{\pm} =
\frac{1}{2}\left[H_{qq} + H_{gg} \pm
\sqrt{(H_{qq} - H_{gg})^2 + 4H_{qg}H_{gq}} \right] .
\end{equation}

Obviously, $\Omega_{+} > \Omega_{-}$.
Eq.~(\ref{g1general})
includes  the coefficient functions
$C_{\pm}(\omega)$ that should be specified. This can be done by different ways.
Below we obtain $C_{\pm}$ in terms of the phenomenological quark and
gluon inputs and then calculate the perturbative contributions to these
inputs.
However before doing so, we first calculate the
anomalous dimensions $H_{ik}$,~ $(i,k = q,g)$.

\section{Calculating $H_{ik}$}

As explicitly shown in eq.~(\ref{omega}),  $\Omega_{\pm}$ are expressed
in terms of the amplitudes $H_{ik}$.
As we are going to  calculate  $\Omega_{\pm}$ to all orders,
we consequently have to know $H_{ik}$ to all orders in $\alpha_s$. To this
aim we write and solve some IREE for $H_{ik}$. The lhs of the new IREE
again corresponds to
differentiating the amplitudes with respect to
$-\mu^2 (\partial/ \partial \mu^2)$ (cf~Eqs.~\ref{system}).
The rhs include, besides the
ladder contributions involving $H_{ik}$, their Born
terms

\begin{equation}
\label{hikborn}
H_{ik}^{Born} =  a_{ik}/\omega,
\end{equation}
and non-ladder contributions $V_{ik}$, which we specify later. The system
of eqs. reads

\begin{eqnarray}
\label{systemh}
\omega H_{qq} &=& a_{qq} +V_{qq} +H_{qq}^2 + H_{qg}H_{gq}, \nonumber \\
\omega H_{gg} &=& a_{gg} +V_{gg} +H_{gg}^2 + H_{gq}H_{qg}, \nonumber \\
\omega H_{qg} &=& a_{qg} +V_{qg} +H_{qg}( H_{qq} + H_{gg}), \nonumber \\
\omega H_{gq} &=& a_{gq} +V_{gq} +H_{gq}( H_{qq} + H_{gg}) ~ .
\end{eqnarray}

The non-ladder DL terms $V_{ik}$
appear in Eqs.~(\ref{systemh}) when the parton with
minimal transverse momentum is a non-ladder
gluon\footnote{We use the Feynman gauge through the paper.}. Such a gluon
can be factorized, i.e. its propagator is attached to external lines.
As the factorized gluon bears a colour quantum number,
the  remaining DL
contribution -defined  $H^c$ for the sake of simplicity - gets also
a coloured content. When the factorized gluon propagates in the $t$ -channel,
 $H^c$ belongs
to the octet $t$-channel representation of $SU(3)$,
whereas all $H_{ik}$ belong to
the singlet representation. Therefore in order to solve
Eqs.~(\ref{systemh}) one has to calculate the octet amplitudes
 first.
Fortunately, they
can be approximated quite well by their Born values as they
fall with energy very rapidly. With this approximation
one obtains

\begin{equation}
\label{vborn}
V_{ik} =\frac{ m_{ik}}{\pi^2}D(\omega) ,
\end{equation}
where
\begin{equation}
\label{m}
m_{qq} =  \frac{C_F}{2N},~
m_{gg} = -2N^2,~
m_{qg} = n_f \frac{N}{2},~
m_{gq} = -N C_F.
\end{equation}
where $n_f$ is the number of the flavours. We assume $n_f = 4$. 
Furthermore $D(\omega)$ in Eq.~(\ref{vborn})
accounts for the running QCD effects for
$V_{ik}$. According to Ref.~\cite{egt1} it is given by

\begin{equation}
\label{d}
D(\omega) = \frac{1}{2b^2} \int_0^{\infty} d \rho e^{- \omega \rho}
\ln \big( (\rho + \eta)/ \eta\big)
\Big[  \frac{\rho + \eta}{(\rho + \eta)^2 + \pi^2} +
\frac{1}{\rho + \eta}\Big]
\end{equation}
where $\eta = \ln(\mu^2/ \Lambda_{QCD}^2)$ and $b = (33 - 2 n_f)/ 12 \pi$.

All $a_{ik}$ in Eqs.~(\ref{systemh}) are proportional to $\alpha_s$.
However, in Ref.~\cite{egt1} it is suggested that in contrast to the DGLAP
prescription
 $\alpha_s = \alpha_s(k^2_{\perp})$ ($k_{\perp}$ is the
transverse momenta of the ladder partons), which holds universally in
every  rung of the ladder, the arguments of
$\alpha_s$  are  different for different kinds of the rungs. In particular,
the expressions for the quark-quark and the gluon-gluon rungs
  include $\alpha_s$
depending on the time-like virtuality of the intermediate gluon,
leading to the following expressions  for $a_{qq}$ and $a_{gg}$:

\begin{equation}
\label{aborndiag}
a_{qq} = \frac{A(\omega)C_F}{2\pi},
~a_{gg} = \frac{2A(\omega) N}{\pi},~
\end{equation}
where

\begin{equation}
\label{a}
A(\omega) = \frac{1}{b} \Big[\frac{\eta}{\eta^2 + \pi^2} -
\int_{0}^{\infty} \frac{d \rho e^{- \omega \rho}}
{(\rho + \eta)^2 + \pi^2}\Big] ~.
\end{equation}
The $\pi^2$-terms  appear in Eq.~(\ref{a}) because $a_{qq}$ and $a_{gg}$
involve $\alpha_s$ with the time-like argument.
On the other hand, in the expressions for the rungs mixing quarks and gluons,
$\alpha_s$ depends the space-like virtuality of
the ladder gluons, which is approximated
by an expression similar to (\ref{a}) without the $\pi^2$ -terms.
So, we arrive at

\begin{equation}
\label{aborn}
a_{gq} = -\frac{n_f A'(\omega)}{2\pi},~
a_{qg} =\frac{ A'(\omega)C_F}{\pi} ~,
\end{equation}
with
\begin{equation}
\label{aprime}
A'(\omega) = \frac{1}{b} \Big[\frac{1}{\eta} -
\int_{0}^{\infty} \frac{d \rho e^{- \omega \rho}}
{(\rho + \eta)^2}\Big] 
~.
\end{equation}

Once $a_{ik}$ and $V_{ik}$ are determined, it is easy to obtain
explicit expressions for $H_{ik}$. Indeed, defining $b_{ik}$ as

\begin{equation}
\label{b}
b_{ik} = a_{ik} + V_{ik} ,
\end{equation}
and

\begin{equation}
\label{Y}
Y = - \sqrt{ \Big(\omega^2 - 2( b_{qq} + b_{gg}) +
\sqrt{\left[(\omega^2 - 2( b_{qq} + b_{gg}))^2  - 4(b_{qq} -b_{gg})^2
-16 b_{qg} b_{gq} \right]} ~ \Big)/2} ~,
\end{equation}
we obtain

\begin{eqnarray}
\label{solh}
H_{gg} &=&\frac{1}{2} \Big(\omega + Y + \frac{ b_{qq} - b_{gg}}{Y}\Big),
H_{qq}= \frac{1}{2} \Big(\omega + Y - \frac{ b_{qq} - b_{gg}}{Y}\Big),
\\ \nonumber
H_{gq} &=& -\frac{b_{gq}}{Y}, ~~~~~~~~H_{qg} = -\frac{b_{qg}}{Y} ~.
\end{eqnarray}
As a matter of fact, the expression (\ref{Y}) for $Y$ appears
in solving Eqs.~(\ref{systemh}) as one of the four
solutions of a biquadratic algebraic
equation.  The signs before the roots in Eq.~(\ref{Y})
are chosen so to get a matching between
$H_{ik} = a_{ik}/ \omega + O(1/\omega^2)$
when $\omega \geq 1$ and the Born contributions of Eqs.~(\ref{aborn}).
Now all ingredients for $\Omega_{\pm}$ are specified and $\Omega_{\pm}$ can
be obtained with numerical calculations.

\section{Intercept of $g_1$}

 Eqs.~(\ref{g1general}) and (\ref{g1primegeneral}) give general expressions
for $g_1$. These expressions  account for both  DL contributions and running
$\alpha_s$ effects. The integrands in  
Eqs.~(\ref{g1general},\ref{g1primegeneral}) contain the coefficient 
functions and exponentials. In the present Sect. 
we consider the exponentials. We will study the coefficient functions in 
the next Sect.     
When the limit $x \to 0$ is considered,  one can neglect the
exponents with $\Omega_-$ in these equations
and simplify the expression
for $\Omega_+$. Indeed,
the behaviour of $g_1$ in this limit is driven by the leading
singularity in Eq.~(\ref{omega}). The singularities of   $\Omega_+$ are
related to the branching points of the square root.
 Using Eq.~(\ref{Y}) one can see that the leading
singularity  of $\Omega_+(\omega)$ is given by the rightmost
root,  $\omega_0$ of the equation below:

\begin{equation}
\label{master}
\omega^4 - 4 (b_{qq} + b_{gg})\omega^2 + 16
(b_{qq} b_{gg} -   b_{qg} b_{gq}) = 0~.
\end{equation}
Therefore,   Eqs.~(\ref{g1general},\ref{g1primegeneral})predict that

\begin{equation}
\label{g1asympt}
g_1 \sim C(\omega_0)(1/x)^{\omega_0}(Q^2/\mu^2)^{\omega_0/2}
\end{equation}
when $x \to 0$, where  $\omega_0$ is discussed below  and 
the factor $C(\omega_0)$ in Sect. V.  
Eq.~(\ref{master})
 can be solved numerically. In our approach, $\omega_0$ depends on
$\eta = \ln(\mu^2/ \Lambda_{QCD}^2)$. The result of the numerical
calculation for $\omega_0 = \omega_0(\eta)$ is represented by the curve 1
in  Fig.~1. This curve first grows with $\eta$,
achieves a maximum where approximately $\omega_0 = 0.86$ and smoothly
decreases for large $\eta$. In other
words we have obtained  that the intercept depends strongly on the
infrared cut-off $\mu$ for small values of  $\mu$ and smoothly thereafter.
Quite a similar situation was occurring in Refs.~\cite{egt2} for
the intercept of the non-singlet structure function $g^{NS}_1$.
We suggest a possible  explanation for this effect.  The cut-off $\mu$ is
defined as the starting point in the description of  the
perturbative  evolution. Everything that affects the
intercept at  scales smaller than $\mu$ is attributed to
non-leading effects and/or non-perturbative
contributions. Had they been accounted for, the intercept would have
been $\mu$-independent. Without those non-leading/non-perturbative effects
taken into account, we then observe an important
 $\mu$-dependence, which becomes weaker for large $\mu$. The maximal
perturbative contribution
to the intercept $\Delta_S$ of $g_1$ is obtained from  the
maximal value of $\omega_0$. Therefore
we estimate the intercept as

\begin{equation}
\label{intercept}
\Delta_S = max (\omega_0(\eta)) \approx 0.86 ~.
\end{equation}

Eq.~(\ref{intercept}) includes the contributions of both virtual quarks
and gluons.
These contributions have opposite signs and partly cancel each other.
It is interesting to note that when only
virtual gluon contributions are taken into account,
this purely gluonic
intercept, $\Delta_g$ is given by the maximum of the curve 2 in
Fig.~1, which is  slightly greater than 1.
This  value
obviously exceeds the unitarity limit, similarly to the intercept
of the LO BFKL\cite{bfkl}, though in a much softer way.
Fortunately, by including also the contributions of the virtual quarks
 the intercept decreases down to $\Delta_S$ of Eq.~(\ref{intercept}),
without violating unitarity.

We would like also to add the following observation.
The total resummation of the DL contributions to $g_1$ performed in
Refs.~\cite{ber, ber1} was used in Refs.~\cite{kw}-\cite{z}, where
$\alpha_s$  was suggested to be fixed at the scale $Q^2$, as in
the framework of DGLAP. In other words, the parametrization
$\alpha_s^{DL}=\alpha_s(Q^2)$ was used in the DL expression
for the intercept,
$\Delta_S^{DL} = 3.5 \sqrt{3 \alpha_s^{DL}/(2 \pi)}$
obtained in  Ref.~\cite{ber}. Now, comparing  this expression
and Eq.~(\ref{intercept}),  one easily concludes that the effective
scale of  $\alpha_s^{DL}$
has no relation
with $Q^2$. This is in a complete accordance with the results of
Refs.~\cite{egt1},\cite{egt2}. Indeed, the parametrisation
$\alpha_s = \alpha_s(Q^2)$ is valid only when the factorisation
of the longitudinal and transverse momentum space is assumed.
This approximation is correct for the
kinematic Al region of large $x$ and does not hold for small $x$.

\section{Coefficient functions $C_{\pm}$}

In order to find $g_q$ and $g_g$ in
Eqs.~(\ref{g1general}, \ref{g1primegeneral}), one has to finally specify
the coefficient functions $C_{\pm}$.
One  way to do it
is to impose the matching

\begin{equation}
\label{match}
g_q(x, Q^2)|_{Q^2 = \mu^2} = \widetilde{\Delta q}(x_0),
~~~~g_g(x, Q^2)|_{Q^2 = \mu^2} =  \widetilde{\Delta g} (x_0)
\end{equation}
where $x_0 = \mu^2/s$; 
$\widetilde{\Delta q}(x_0)$ and $\widetilde{\Delta g} (x_0)$ 
are
the initial densities of the polarized quarks and gluons respectively.
Using the integral transform of Eqs.~(\ref{g1general},\ref{g1primegeneral})
at $Q^2 = \mu^2$
in the $\omega$-space the matching of
 Eq.~(\ref{match}) can be rewritten as

\begin{equation}
\label{matchomega}
C_+ + C_- = \Delta q,
~~~~C_+\frac{X + \sqrt{R}}{2H_{qg}} +
C_- \frac{X - \sqrt{R}}{2H_{qg}} = \Delta g ~ ,
\end{equation}
with both $\Delta q$ and  $\Delta g$ depending on $\omega$.
Combining Eqs.~(\ref{g1general}, \ref{g1primegeneral}) and
(\ref{matchomega}), we express $C_{\pm}$ through
$\Delta q, \Delta g$     and arrive at

\begin{equation}
\label{g1}
g_q(x, Q^2) = \int_{- \imath \infty}^{\imath \infty}
\frac{d \omega}{2 \pi \imath} (1/ x)^{\omega}
\Big[ \Big(  A^{(-)} \Delta q +
B \Delta g \Big)  e^{\Omega_+ y} +
 \Big( A^{(+)} \Delta q - B \Delta g \big) e^{\Omega_- y} \big]~,
\end{equation}

\begin{equation}
\label{g1prime}
g_g(x, Q^2) = \int_{- \imath \infty}^{\imath \infty}
\frac{d \omega}{2 \pi \imath} (1/ x)^{\omega}
\Big[\Big( E \Delta q + A^{(+)}\Delta g
\Big) e^{\Omega_+ y} +
\Big(- E\Delta q +  A^{(-)}\Delta g \Big)
 e^{\Omega_- y}\Big]
\end{equation}
with
\begin{equation}
\label{abe}
A^{(\pm)} = \Big(\frac{1}{2} \pm \frac{X}{2 \sqrt{R}}\Big),~
B = \frac{H_{qg}}{\sqrt{R}},~
E = \frac{H_{gq}}{\sqrt{R}} ~.
\end{equation}
Eqs.~(\ref{g1}) and (\ref{g1prime}) describe the $Q^2$ evolution of
the polarized quark and gluon densities $g_q$ and $g_g$
from $Q^2 = \mu^2$,  where they are
$\tilde{\Delta} q(\mu^2/s)$ and $\tilde{\Delta} g(\mu^2/s)$ respectively,
to the region $Q^2 \gg \mu^2$.
In the framework of our approach, $\mu^2$ is the starting point for
the $Q^2$ -evolution. Then one option would be to fix the initial parton
densities
$\tilde{\Delta} q$ and $\tilde{\Delta} g$  
from phenomenological considerations.
 
In this case, one can use 
Eqs.~(\ref{g1},\ref{g1prime},\ref{abe}) in order to fix 
the sign of $g_1$ in the small-$x$ region. 
The problem of the sign involves an interplay of the gluon and  quark 
contributions to $g_1$ (see e.g. Refs.~\cite{ber},\cite{r}). 
Let us estimate  the sign of $g_1$ in 
the small-$x$ region, calculating the asymptotics of   
$A^{(\pm)},~B,~E$.  
When $x \to 0$, the main contributions in
Eqs.~(\ref{g1},\ref{g1prime}) come from the terms proportional to
$\exp[\Omega_+ y]$. According to the results of Sect.~4, the small-$x$ 
asymptotics of $g_1$ is 

\begin{equation}
\label{asgeneral}
g_1 \sim 
[(A^{(-)}_S + E_S) \Delta q + (A^{(+)}_S + B_S) \Delta g ]~ 
(1/x)^{\Delta_S} (Q^2/\mu^2)^{\Delta_S/2}
\end{equation} 
where we have provided the factors $A^{(\pm)}, B, E$ with the subscript 
$``S''$ in order to show their explicit dependence 
on $\omega = \Delta_S$. Substituting 
the numerical values 
$A^{(-)}_S = -0.31,A^{(+)}_S = 1.31, B_S = 0.52, E_S = -0.79$ 
into Eq.~(\ref{asgeneral}), we arrive at  

\begin{equation}
\label{as}
g_1 \sim 
[- 1.1 \Delta q + 1.8 \Delta g ]~
(1/x)^{\Delta_S} (Q^2/\mu^2)^{\Delta_S/2}
 ~. 
\end{equation} 
As $\Delta q$ is positive,  $g_1$ is negative when 
\begin{equation}
\label{deltaas}
\Delta q > 1.7 \Delta g ~.
\end{equation}
 
Eqs.~(\ref{as},\ref{deltaas}) are expressed in terms of 
 the quark and 
gluon initial densities  $\tilde{\Delta q},~\tilde{\Delta g}$ 
defined (see Eq.~(\ref{match})) at the 
scale $x \approx \mu^2/s$. 
On the contrary, the standard DGLAP expressions for $g_1$ involve 
 the initial densities  $\tilde{\delta} q$ and $\tilde{\delta} g$ 
defined at the scale $x \sim 1$. 
In order to compare our results to DGLAP, we should express 
  $\tilde{\Delta} q$ and $\tilde{\Delta} g$ in terms of 
 $\tilde{\delta} q$ and $\tilde{\delta} g$. 
In the $\omega$-space, it means expressing 
 $\Delta q(\omega)$ and $\Delta g(\omega)$ 
through $\delta q(\omega)$ and $\delta g(\omega)$. 
We can do it within our framework by   
using the evolution of   $\tilde{\Delta} q$ and $\tilde{\Delta} g$ 
with respect to $s$ at fixed $Q^2 \approx \mu^2$ 
from the starting point $s \approx \mu^2$ to the region $s \gg \mu^2$. 
The system of the IREE for  
 $\Delta q$ and $\Delta g$  
is similar to Eq.~(\ref{system}) with the exeption of  
the fact that $\tilde{\Delta} q$ and $\tilde{\Delta} g$ do not  
depend on $Q^2$ and therefore the 
IREE for them do not include the derivatives  
$\partial/\partial y$.    
Therefore,  we arrive at the following algebraic IREE:    

\begin{eqnarray}
\label{systeminput}
  \Delta q (\omega) &=& (e^2_q/2) \delta q(\omega) + (1 /\omega)
\left[H_{qq}(\omega) \Delta q(\omega)  +
H_{qg}(\omega) \Delta g (\omega) \right]~, \nonumber \\
 \Delta g(\omega) &=& (e^2_q/2) \hat{\delta g}(\omega) + (1/\omega)
\left[H_{gq}(\omega) \Delta q(\omega)  +
H_{gg}(\omega) \Delta g(\omega) \right] ~. 
\end{eqnarray}
 where $e^2_q$ is the sum of the quark electric 
charges ($e^2_q = 10/9$ for $n_f = 4$),  $\delta q$ is the sum of the 
initial quark and antiquark densities 
 and        
$\hat{\delta g} \equiv  -  (A'(\omega)/2 \pi \omega^2) \delta g$ 
($A'(\omega)$ is defined in  Eq.~(\ref{aprime}))  
is the starting 
point of the evolution of the gluon density $\delta g$.       
Eqs.~(\ref{systeminput}) describe the $s$-evolution of the quark and gluon 
inputs from $s \sim \mu^2$ to $s \gg \mu^2$. 
Solving Eqs.~(\ref{systeminput}), we
obtain:

\begin{equation}
\label{inputq}
\Delta q= (e^2_q/2)
\frac{ 
\big[\omega (\omega -H_{gg}) \delta q + \omega H_{qg}\hat{\delta g}\big] }
{\big[\omega^2 - \omega(H_{qq} + H_{gg}) + (H_{qq}H_{gg} -
H_{qg}H_{gq})\big]}~,
\end{equation}
\begin{equation}
\label{inputg}
\Delta g= (e^2_q/2)
 \frac{ 
\big[\omega H_{gq} \delta q + \omega(\omega - H_{qq})\hat{\delta} g \big]}
{\big[\omega^2 - \omega(H_{qq} + H_{gg}) + (H_{qq}H_{gg} -
H_{qg}H_{gq})\big]}  ~.
\end{equation}

Substituting
Eqs.~(\ref{inputq},\ref{inputg}) into Eq.~(\ref{as}) allows 
to express asymptotics of   $g_1$ in 
terms of $\delta q, \delta g$: 
\begin{equation}
\label{newas}
g_1 \sim (1/2)  
[ -1.2 \delta q - 0.08 \delta g~ ]~
(1/x)^{\Delta_S} (Q^2/\mu^2)^{\Delta_S/2} ~. 
\end{equation} 
   
In contrast to Eq.~(\ref{as}) which involves  the parton distributions 
  $\tilde{\Delta} q$ and $\tilde{\Delta} g$ defined at the low-$x$  
scale, at $x \approx \mu^2/s$,  
Eq.~(\ref{newas}) is expressed in terms of the densities  
$\tilde{\delta} q$ and $\tilde{\delta} g$ at $x \sim 1$. These 
densities can be 
fixed from phenomenological considerations.   
Eq.~(\ref{newas}) states that $g_1$ can be positive at $x \ll 1$ 
only when  $\delta g$ 
is negative and large:  
 
\begin{equation}
\label{newdeltaas}
 15 \delta q\ + \delta g < 0 ~. 
\end{equation}
 

\section{Summary and outlook}

By constructing and solving the system (\ref{system}) of infrared
evolution equations,
we have obtained the explicit expressions
(\ref{g1},\ref{g1prime}) for the polarized quark and gluon distributions
$g_q(x, Q^2)$ and $g_g(x, Q^2)$. These expressions account for DL
contributions to all orders and, at the same time, for the
running $\alpha_s$ effects. Both  $g_q(x, Q^2)$ and $g_g(x, Q^2)$
depend on the low-$x$ quark and  gluon distributions, $\Delta q$ and
$\Delta g$. Eq.~(\ref{as}) 
shows that $g_1$ has the negative sign in the small-$x$ region if 
$\Delta q > 1.7 \Delta g$. 
By the same method, and using the $s$-evolution 
at $Q^2 \approx \mu^2$,    
the distributions  $\Delta q$ and $\Delta g$ are 
expressed  in  Eq.~(\ref{inputq},\ref{inputg}) in terms of the initial 
densities  $\delta q$ and $\delta g$ 
which are supposed to be fixed
from phenomenological considerations at $x \sim 1$.  
Eq.~(\ref{newas}) shows that 
$g_1$ is positive when the initial gluon density is negative and large: 
 $\delta g < - 15 \delta g$, otherwise $g_1$ is negative.    

We obtain that the expressions~(\ref{g1},\ref{g1prime}) lead 
to
the Regge behaviour (\ref{as},\ref{newas}) of
$g_1$ when $x \to 0$.  
The value of the intercept $\Delta_S$ depends on the infrared cut-off
$\mu$, but this dependence is quite weak for $\mu$ much larger than
$\Lambda_{QCD}$.
The estimated value of the intercept is given by
Eq.~(\ref{intercept}), and is a factor of $2.2$ larger than the non-singlet
intercept. The value of  $\Delta_S = 0.86$ is in a good agreement with the 
estimate $\Delta_S = 0.88$ obtained in Ref.~\cite{koch} 
from analysis of the HERMES data.  
Our results also show that
accounting for the gluon contributions only one would obtain a value of
the intercept exceeding unity and therefore violating unitarity
- similarly to the LO BFKL intercept -  whereas the inclusion of the quark
contributions stabilises the result. We prove that it is unrealistic  to
combine the resummation of the DL contributions to $g_1$ with
the DGLAP-like parametrisation for  $\alpha_s$ in the expressions
for the intercepts.

Finally, we note that it would be very interesting
to implement our results by non-perturbative (lattice) calculations in
order to check explicitly the independence of the total
intercept on $\mu$.

\section{Acknowledgement}

We are grateful to S.I.~Krivonos and S.M.~Oliveira
for useful discussions concerning the
numerical calculations.
The work is supported by grants POCTI/FNU/49523/2002
and RSGSS-1124.2003.2 .

\section{Figure captions}

Fig.~1: Dependence on $\eta$ of the rightmost root of Eq.~(\ref{master}),
 $\omega_0$.
Curve~2 corresponds to the case
when gluon contributions only are taken into account; curve~1 is the result of
accounting for both gluon and quark contributions.




\begin{thebibliography}{99}

\bibitem{ar} M.A.~Ahmed and G.G.Ross. Nucl. Phys.B 111(1976)441.

\bibitem{ap} G.~Altarelli and G.~Parisi. Nucl.Phys.B 126(1977) 298.

\bibitem{dgl}
 V.N.~Gribov and L.N.~Lipatov. Sov.J.Nucl.Phys.15(1978) 438 and 675;
  Yu.L.~Dokshitzer.  Sov.Phys.JETP 46(1977)641.

\bibitem{a} G.~Altarelli, R.~Ball, S.~Forte and G.~Ridolfi.
Acta Phys.Polon.B29(1998)1201; Nucl. Phys. B496(1999)337. 

\bibitem{b}  J.~Blumlein and H.~Bottcher. Nucl. Phys.B 636(2002)225.

\bibitem{grsv}  M.G.~Gluck, E.~Reya, M.~Straumann and W.~Vogelsang.
Phys. Rev. D53(1996)4775.

\bibitem{ber1} B.I.~Ermolaev, S.I.~Manaenkov and M.G.~Ryskin.
Z.Phys.C69(1996)259; ~J.~Bartels, B.I.~Ermolaev and M.G.~Ryskin.
Z.Phys.C 70(1996)273.

\bibitem{ber} J.~Bartels, B.I.~Ermolaev and M.G.~Ryskin.
Z.Phys.C 72(1996)627.

\bibitem{abcmv} D.~Amati, A.~Bassetto, M.~Ciafaloni, G.~Marchesini and
G.~Veneziano. Nucl.Phys.B 173(1980)428.

\bibitem{kw} J.~Kwiecinski. Acta.Phys.Polon.B 29(2001)1201;

\bibitem{kk} D.~Kotlorz and A.~Kotlorz. Acta.Phys.Polon.B 32(2001) 2883;

\bibitem{bbk}B.~Badalek, J.~Kiryluk and J.~Kwiecinski.
Phys.Rev.D 61(2000)014009.

\bibitem{kz}J.~Kwiecinski and B.~Ziaja. Phys.Rev.D 60(1999)9802386;
 J.~Kwiecinski and B.~Zaija. hep-ph/9802386.

\bibitem{z}B.~Ziaja. Phys. Rev. D66(2002)114017; hep-ph/0304268.

\bibitem{egt1} B.I.~Ermolaev, M.~Greco and S.I.~Troyan.
Phys.Lett.B 522(2001)57.



\bibitem{egt2} B.I.~Ermolaev, M.~Greco and S.I.~Troyan.
Nucl.Phys.B 594 (2001)71; ibid 571(2000)137.

\bibitem{bfkl}  V.S.~Fadin, E.A.~Kuraev and L.N.~Lipatov.
Sov.Fiz.JETP, 44(1976)443, ibid 45(1977)199.

\bibitem{kat} J.~Soffer and O.V.~Teryaev. Phys.Rev.D 56(1997)1549;
A.L.~Kataev, G.~Parente, A.V.Sidorov. CERN-TH-2001-058;
Phys.Part.Nucl.34(2003)20; Fiz.Elem.Chast.Atom.Yadra 34(2003)43;
Nucl.Phys.A666(2000)184.
A.V.~Kotikov, A.V.~Lipatov, G.~Parente, N.P.~Zotov, Eur.Phys.J.C26(2002)51;
V.G.~Krivokhijine, A.V.~Kotikov, hep-ph/0108224;
 A.V.~Kotikov, D.V~. Peshekhonov,  hep-ph/0110229.

\bibitem{r} A.~De~Roeck, A.~Deshpande, V.W.~Hughes, J.~Lichtenstadt,
G.~Radel. Eur.Phys.J.C6(1999)121.

\bibitem{koch}   N.I.Kochelev, K.Lipka, W.-D.Nowak, V.Vento, A.V.Vinnikov. 
Phys.Rev. D67 (2003) 074014

\end{thebibliography}
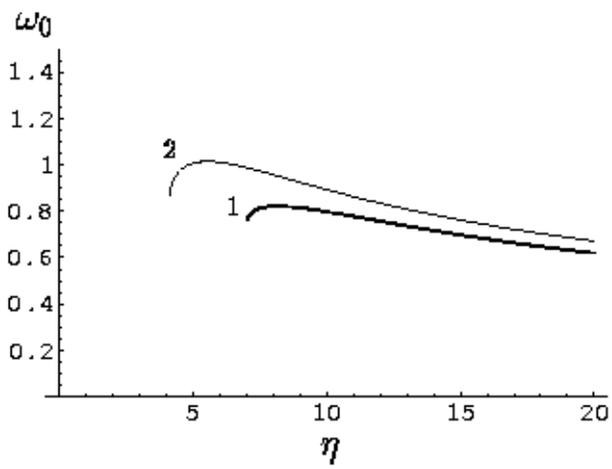
\end{document}